\documentclass[aps,prl,10pt,twocolumn,superscriptaddress,notitlepage,floatfix]{revtex4-1}
\usepackage[utf8]{inputenc}
\usepackage{amsmath,amsfonts,amssymb,color,graphicx,xcolor,physics}
\usepackage[colorlinks]{hyperref}
\usepackage{dcolumn}%

\definecolor{rdvcolor}{rgb}{0,0.5,0}

\begin{document}
\title{Distributed quantum error correction for chip-level catastrophic errors}

\author{Qian Xu} 
\thanks{These authors contributed equally.}
\affiliation{Pritzker School of Molecular Engineering, The University of Chicago, Chicago IL 60637, USA}

\author{Alireza Seif} 
\thanks{These authors contributed equally.}
\affiliation{Pritzker School of Molecular Engineering, The University of Chicago, Chicago IL 60637, USA}

\author{Haoxiong Yan}
\affiliation{Pritzker School of Molecular Engineering, The University of Chicago, Chicago IL 60637, USA}

\author{Nam Mannucci}
\affiliation{Pritzker School of Molecular Engineering, The University of Chicago, Chicago IL 60637, USA}

\author{Bernard~Ousmane~Sane}
\affiliation{Graduate School of Media and Governance, Keio University, 5322 Endo, Fujisawa 252-0882, Japan}

\author{Rodney Van Meter}
\affiliation{Faculty of Environment and Information Studies, Keio University, 5322 Endo, Fujisawa 252-0882, Japan}

\author{Andrew N. Cleland}
\affiliation{Pritzker School of Molecular Engineering, The University of Chicago, Chicago IL 60637, USA}
\affiliation{Center for Molecular Engineering and Material Science Division, Argonne National Laboratory, Lemont IL 60439, USA}

\author{Liang Jiang} 
\email{liang.jiang@uchicago.edu}
\affiliation{Pritzker School of Molecular Engineering, The University of Chicago, Chicago IL 60637, USA}
%\affiliation{AWS Center for Quantum Computing, Pasadena, CA 91125, USA}

\date{\today}
\begin{abstract}
Quantum error correction holds the key to scaling up quantum computers. Cosmic ray events severely impact the operation of a quantum computer by causing chip-level catastrophic errors, essentially erasing the information encoded in a chip. Here, we present a distributed error correction scheme to combat the devastating effect of such events by introducing an additional layer of quantum erasure error correcting code across separate chips. We show that our scheme is fault tolerant against chip-level catastrophic errors and discuss its experimental implementation using superconducting qubits with microwave links. Our analysis shows that in state-of-the-art experiments, it is possible to suppress the rate of these errors from 1 per 10 seconds to less than 1 per month.
\end{abstract}
\maketitle

{\it Introduction-- } Extreme sensitivity to external noise is one of the main obstacles in building and operating large-scale quantum devices. Quantum error correction (QEC) solves this issue by encoding quantum information in a larger space so that the errors can be detected and corrected~\cite{nielsen_chuang_2010,lidar_brun_2013}. For most QEC schemes, the errors need to be small and independent. Existing QEC schemes mostly focus on local and uncorrelated error (or errors with finite-range correlations), see e.g.~\cite{Clemens2004quantum,Nickerson2019analysingcorrelated}.
Long-range correlations, however, can appear if the system is coupled to a common environment, e.g. a bosonic bath~\cite{Alicki2002Dynamical,averin2003active,klesse2005correlated} can negatively impact the performance of QEC~\cite{Aliferis2006QuantumAcc,Aharanov2006Fault}.

Recently, it has been shown that a cosmic ray event (CRE) can cause catastrophic errors by destroying the qubit coherence throughout the superconducting quantum chip for thousands of operation cycles~\cite{vepsalainen2020impact,cardani2021reducing,mcewen2021resolving,mcewen2021resolving}. Upon impact of high-energy rays, phonons are created and spread in the substrate. These phonons then create quasiparticles in the superconducting material, which subsequently induces qubit decay~\cite{mcewen2021resolving}. Even though these events are rare, their effect is devastating as they cause fast correlated relaxation ($T_1$ error) in all the qubits in a chip that essentially erases the encoded quantum information~\cite{mcewen2021resolving}, which is especially detrimental to long computational tasks that could take several hours~\cite{gidney2021factor}. Moreover, the adverse effect of CREs is not limited to superconducting qubits. Semiconductor spin qubits~\cite{spinqubits} and qubits based on Majorana fermions~\cite{rainis2012majorana,martinis2021saving} also suffer from the charge noise and quasiparticle poisoning that are resulted from CRE, respectively. One approach to reducing the impact of CREs is through changing the design of the device, for example, by introducing phonon and quasiparticle traps~\cite{nsanzineza2014trapping,patel2017phonon,henriques2019phonon} and enhancing phonon relaxation in the device~\cite{martinis2021saving}. Such an approach requires a great deal of engineering, with details depending on the specific platform of interest.

\begin{figure}
    \centering
    \includegraphics[width=\columnwidth]{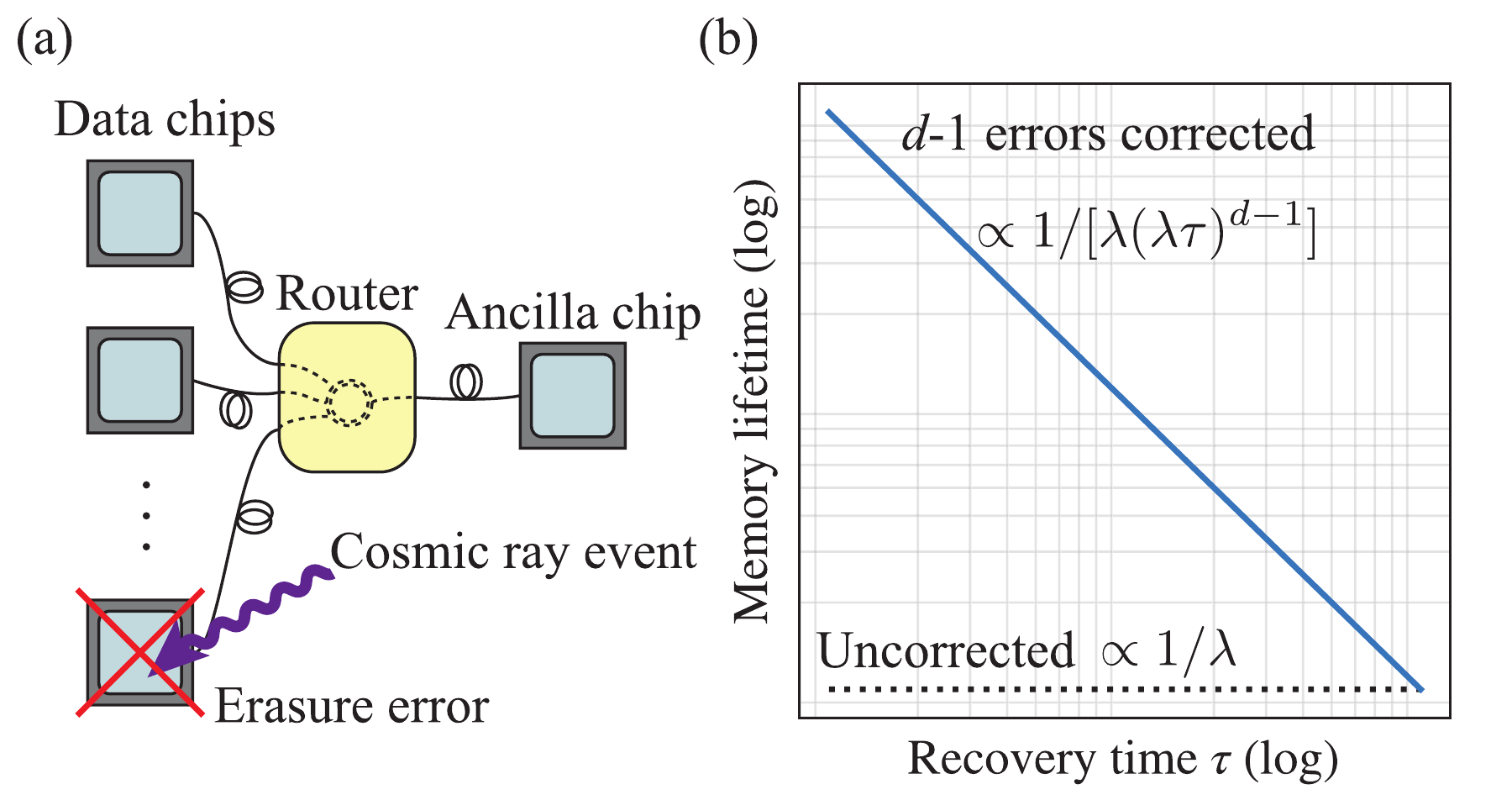}
    \caption{(a) Schematic of the distributed encoding setup. Information is encoded in an error correcting code that is distributed across multiple data chips, which are coupled to an ancilla chip using microwave links and a router. 
    % A cosmic ray event leads to an erasure error on a single chip. The  error can be corrected fault-tolerantly  by applying ancilla-assisted syndrome measurements.
    The CRE-induced erasure errors on the data chips can be corrected by applying ancilla-assisted syndrome measurements.
    (b) Using a code that can correct $d-1$ erasure errors, we can suppress the rate of the CRE-induced catastrophic events to $\propto\lambda(\lambda \tau)^{d-1}$, where $\lambda$ is the CRE rate and $\tau$ is the duration of error recovery operation.}
    \label{fig:schematic}
\end{figure}

In this work, we take a different approach and use a distributed error correcting scheme to detect and correct correlated errors by CREs. Distributed hardware architectures, connecting smaller nodes into a tightly-coupled system using an interconnect network, have been proposed to achieve scalability for a single computation~\cite{Duan04,jiang07:PhysRevA.76.062323,kim09:_integ_optic_ion_trap,lim05:_repeat_until_success,nickerson2013topological,oi06:_dist-ion-trap-qec,van-meter10:dist_arch_ijqi,van-meter16:_ieee-comp,Monroe14}.  Here, we repurpose these architectures to improve fault tolerance. Our approach is system independent and works as long as a quantum network can be built to share entanglement between separate chips. Since CREs are independent stochastic processes~\cite{mcewen2021resolving}, there is no correlation between CREs at different chips, when the interconnects are switched off. In a network of chips, a CRE erases information from one chip, but as we show this event and the specific impacted chip can be detected (see Fig.~\ref{fig:schematic}a). Since the location of the error is now known, we can use erasure QEC to correct the errors and recover the information~\cite{grasel1997codes, knill2005scalable, knill00:loqc-threshold, silva2004erasure}.
% with a lower overhead compared to general error channels. 
We present a low-overhead erasure QEC scheme that is fault tolerant against the CREs and discuss its implementation using superconducting chips connected with microwave links (see e.g., Refs.~\cite{Kurpiers2018, Axline2018,zhong2021deterministic,Gold_2021}), and provide logical-error estimates in state-of-the-art experimental systems. Our analysis indicates that under reasonable assumptions, we are able to suppress the damage from these catastrophic events to higher order and reduce the CRE-induced logical error rate  from 1 every 10 seconds in Ref.~\cite{mcewen2021resolving} to less than 1 per month. We emphasize that while our estimations of code performance are done for a specific platform, our scheme is  general and can be applied to other quantum computing platforms that are severely affected by CREs.

{\it Setup.-- } 
We consider two levels of encoding on $n$ chips, each containing hundred of qubits. The first level uses an error correcting code (e.g., a $10\times10$ surface code~\cite{fowler2012surface}) to protect the information in each chip. In the second level of encoding, we concatenate this code with a $[[n,1,d]]$ QEC code capable of correcting $d-1$ erasure errors~\cite{grasel1997codes}. The operations in the first level should be protected by the surface code.  Therefore, operations in that level are followed by syndrome checks at every step. Upon a CRE on a specific chip, most syndromes of the first-level encoding in that chip will show an error, which reveals the location of the erasure error in the second level. This will subsequently trigger error correction in the second level. We expect that by correcting $d-1$ errors we would be able to suppress the rate of catastrophic events to $\propto\lambda (\lambda \tau)^{d-1}$, where $\lambda$ is the CRE rate in a chip and $\tau$ is the time that it takes for the second-level error correction cycle (see Fig.~\ref{fig:schematic}b). 

\begin{figure*}[t]
    \centering
    \includegraphics[width=2.0\columnwidth]{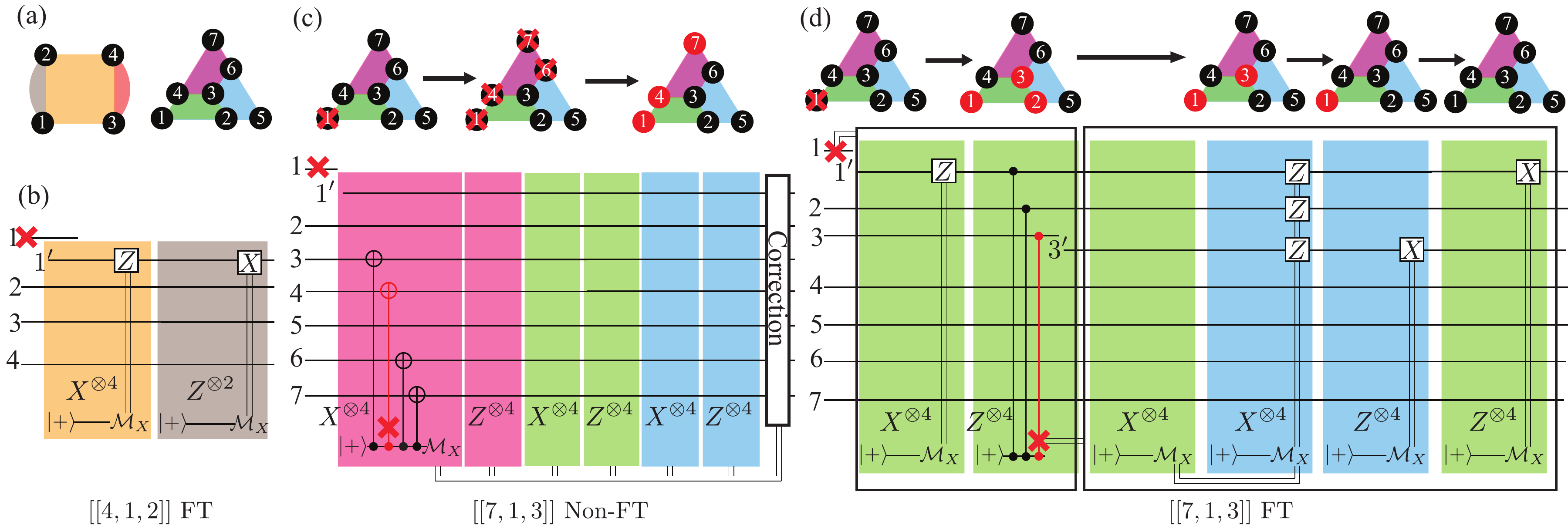}
    \caption{ QEC circuits for correcting erasure errors. (a). The illustration of the [[4,1,2]] (left) and the [[7,1,3]] (right) code. The colored plaquettes represent stabilizer generators that have supports on the surrounding vertices (data qubits). (b) The FT circuit for the [[4,1,2]] code correcting one data erasure (indicated by the red cross). Colored boxes represent an ancilla-assisted measurement of a stabilizer associated with a plaquette of the same color.  The ancilla is initialized in the $|+\rangle$ state, a sequence of $CX/CZ$ gates between the ancilla and the data qubits are applied, and the ancilla is measured in the Pauli $X$ basis. For simplicity, we do not show the gates in the boxes.  (c) The non-FT circuit for the [[7,1,3]] code that is \textit{non-adaptive}. An initial erasure error on a data qubit triggers the circuit, which measures all the six stabilizers in a fixed sequence (pink-green-blue) and applies the correction in the end. We explicitly show the $CX$ gates in the first box (the measurement of $X_3 X_4 X_6 X_7$ stabilizer) to illustrate an erasure error that propagates to multiple data errors and causes a logical failure. As such, the non-FT circuit cannot correct certain consecutive double erasures. On the top, we show the evolution of the data errors for the example trajectory. The red circles indicate qubits with potential Pauli errors (converted from the erasure errors). (d). The FT circuit for the [[7,1,3]] corrects the errors \textit{adaptively}. Suppose another erasure happens during the $CZ$ gate (shown in red) between the ancilla and the third qubit while measuring $Z_1 Z_2 Z_3 Z_4$. Upon detection of this error we stop the stabilizer measurement, discard and replace the ancilla and the thrid qubits and update the erasure-flag error set $\mathcal{E}$ to be $\mathcal{E} = \left\{I, X_{1}\right\} \times\left\{I, P_{3}\right\} \times\left\{I, Z_{1} Z_{2}\right\}$, where $P_3$ indicates an arbitrary Pauli error on the third qubit and the correlated $Z_1 Z_2$ error results from discarding the ancilla that is already entangled with the first and the second qubits. We then measure the stabilizers $X_1 X_2 X_3 X_4$, $X_2 X_3 X_5 X_6$, $Z_2 Z_3 Z_5 Z_6$, $Z_1 Z_2 Z_3 Z_4$ to correct the possible errors within $\mathcal{E}$. }
    \label{fig:FT_QEC_circuits}
\end{figure*}

For example, we can use the [[4,1,2]] code~\cite{grasel1997codes} to correct single erasure errors. As shown in Fig.~\ref{fig:FT_QEC_circuits}b, a single CRE event will trigger the QEC circuit to correct the erasure error and successfully restore the original encoding. However, if there is a second CRE erasure event during the erasure correction, the QEC circuit will fail to restore the encoded information, leading to a CRE-induced logical error rate proportional to $\lambda^2 \tau$, which is already suppressed to the second order in $\lambda$. Note that the QEC for $[[4,1,2]]$ is relatively simple because we only care about correcting single CRE errors and do not worry about CRE errors during the QEC operation. In order to use larger-distance codes, e.g., the $[[7,1,3]]$ code~\cite{steane1996multiple},  to suppress the CRE errors to higher orders it is crucial to design the QEC circuit \emph{fault tolerantly} so that all possible relevant CRE events during the QEC should not damage the encoded information.

{\it Fault tolerant error correction for erasure errors.--} 
We assume that by using sufficiently large surface codes in the first level, Pauli error rates due to the failure of the surface QEC are much lower than the rate of the CREs. As such, we only consider the errors induced by the CREs. For simplicity, we assume that a CRE-induced erasure error could propagate through a two-qubit gate and completely erase both involved qubits \footnote{In practice, the propagation of erasure errors at the surface-code level through the gates depends on both the microscopic detail of the erasures and the implementation of the gates. Here we consider the worst-case scenario, where a single erasure can completely erase all the qubits involved in a multi-qubit gate.}. Upon detecting erasure errors on a chip, we replace the erased chip with a chip held in reserve for this purpose. The data qubits on the new chip are randomly initialized. Hence, their erasure errors are converted to detected Pauli errors randomly drawn from $\{I, X, Y, Z\}$ after the chip replacement. The weight of an error is assigned by counting the number of qubits on which the error has non-identity support (including the erasure). We propose a novel fault-tolerant QEC (FTQEC) scheme, which we call the erasure-flag scheme, that satisfies the fault-tolerant criteria~\cite{Aliferis2006QuantumAcc, gottesman2010introduction, chamberland2018flag} (see also Supplementary Material~\cite{SM}). The scheme \textit{adaptively} performs non-destructive stabilizer measurements using one ancilla qubit on an ancilla chip (see Fig.~\ref{fig:schematic}a). A single erasure error that occurs on the ancilla could possibly propagate into multiple data erasures on different data chips. We define such errors as bad errors. However, since we can detect such errors immediately, we get extra information about when and where the errors occur.  So similar to the flag FTQEC for generic Pauli errors~\cite{chao2018quantum, chao2018fault, chamberland2018flag}, the access to the extra information enables us to design protocols that use minimal resources to tolerate the bad errors. In our context, the extra information comes directly from the first-level QEC and does not require additional resources, e.g. flag qubits, in the second level.

The erasure-flag FTQEC protocol using a distance-$d$ code is implemented as follows. (i). Upon detecting erasure errors on the data qubits, replace the erased data qubits (chips), initialize the erasure-flag error set $\mathcal{E}$ which contains the detected data errors, set $s = 0$ which counts the number of bad erasure errors that happen during the protocol and apply the following erasure-QEC. (ii). Measure a set of stabilizers of minimal size 
% (\new{those with  support that overlaps with the support of $\mathcal{E}$}) 
that can be used to correct the current $\mathcal{E}$. (A). If there are $s_{\textrm{new}}$ bad erasures detected in the middle of a stabilizer measurement with $s + s_{\textrm{new}} \leq t$, stop the measurement immediately, update $s$ by adding $s_{\textit{new}}$, replace the erased qubits (chips), update $\mathcal{E}$, and restart (ii). (B). Otherwise, apply a correction in $\mathcal{E}$ based on the measured syndromes. 

The fault tolerance of the protocol is guaranteed by the following two key ingredients. (a). Bad erasures can be immediately detected so that we can keep track of the erasure-flag error set resulting from the bad errors. (b). The erasure-flag error set is correctable (different errors either have different syndromes or differ by a stabilizer) if there are fewer than $t$ faults. We note that similar to the case of the flag-QEC, the second ingredient cannot always be satisfied, and it depends on the codes and syndrome extraction circuits in general. Here we show that the erasure-flag scheme can be applied to the four-qubit and seven-qubit codes using proper QEC circuits, and in Supplementary Material~\cite{SM} we show that it can be applied, more generally, to other codes including the topological surface codes with arbitrary distance.
We show the QEC circuits for the four-qubit and seven-qubit codes in Fig.~\ref{fig:FT_QEC_circuits}. The FT circuit for the [[4,1,2]] code (Fig.~\ref{fig:FT_QEC_circuits}b) corrects a single data erasure at the input. A non-FT circuit for the [[7,1,3]] code (Fig.~\ref{fig:FT_QEC_circuits}c) is triggered by a data erasure error at the input and \textit{non-adaptively} measures a full set of stabilizers in a fixed sequence.
However, an extra erasure that occurs on the ancilla chip during a stabilizer measurement could propagate into multiple data errors and cause a logical failure. As such, the non-FT circuit fails to correct some consecutive double erasures. In contrast, the \textit{adaptive} FT circuit (Fig.~\ref{fig:FT_QEC_circuits}d), which keeps track of the possible error set and measures only a minimal set of stabilizers, can tolerate up to two consecutive erasures on arbitrary qubits.   

\begin{figure}[t]
    \centering
    \includegraphics[width=\columnwidth]{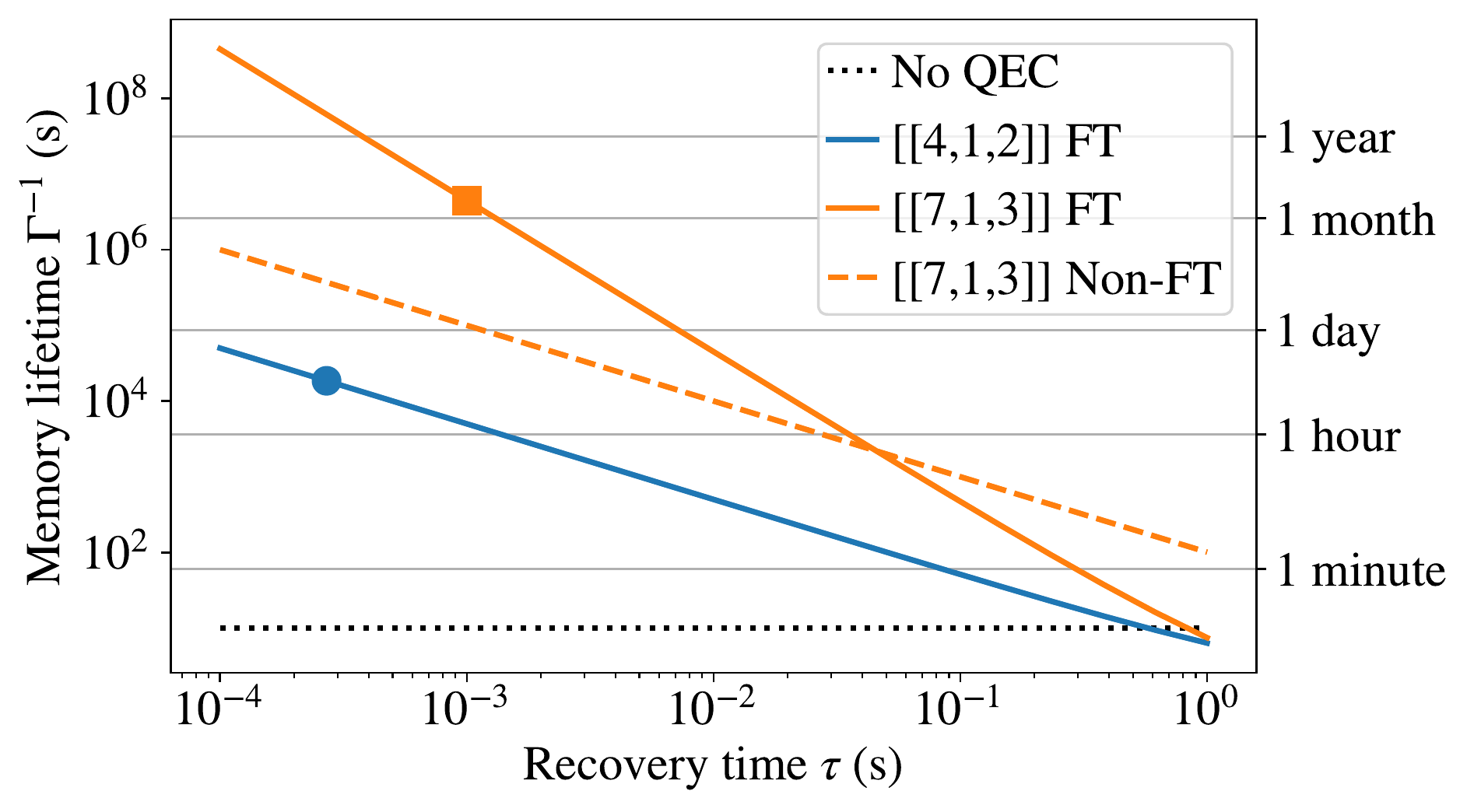}
    \caption{ The estimated rate of the catastrophic events $\Gamma$ with and without error correction. The solid lines show the lower bound of the lifetime with the fault-tolerant implementation, whereas the dashed line is an upper bound of the lifetime obtained without fault-tolerant implementation. The dotted line shows the expected lifetime $\lambda^{-1}$ without error correction. The circle and square markers show our estimate of the improved lifetime for maximal recovery time using experimentally feasible parameters for the $[[4,1,2]]$ and $[[7,1,3]]$ codes, respectively. }
    \label{fig:rates}
\end{figure}

{\it Analysis of the erasure error rates.--} Following Ref.~\cite{mcewen2021resolving}, we model CREs on each chip by a Poisson process $N(t)$, such that $P[N(t)=k]=(\lambda t)^k/k! \exp(-\lambda t)$, where $\lambda$ is the rate of the events whose numerical value is reportedly $1/\lambda = 10\ {\rm{s}}$. Of course, the exact numerical value of $\lambda$ depends on the geometry and other specifications of the chip. However, for simplicity, we assume that this rate can be applied to our setup of interest~\cite{martinis2021saving}. Since the events in each chip are independent, the introduction of additional chips increases the overall rate of the events in the system linearly. Using the FT implementation of a QEC code that corrects $d-1$ erasure errors in a cycle, a catastrophic event might occur if there are more than $d-2$ additional events during the recovery time, $\tau$, following the first event that triggers error correction. Such a catastrophic event leads to a logical failure at the second level of encoding. The rate of these catastrophic events is obtained by taking the product of the rate of the CREs that trigger error correction and the probability that  more than $d-2$ CREs happen in time $\tau$ following the first CRE. For a code over $n$ chips, the former is $n\lambda$. However, since we need an ancilla chip for our QEC scheme, the latter factor should be calculated using the rate $(n+1)\lambda$. Therefore, we find the rate of the catastrophic events, $\Gamma = n\lambda\{1-\exp[-(n+1)\lambda \tau]\sum_{k=0}^{d-2} [(n+1)\lambda \tau]^k/k!\}$. For $n\lambda \tau \ll 1$, we can approximate this  by $\Gamma\approx n\lambda [(n+1)\lambda \tau]^{d-1}/(d-1)!$, which shows the desired error suppression in this regime.  Note that here we considered the worst-case scenario, but not all weight-$d$ (or higher) errors are catastrophic, and some are still correctable. Therefore, by considering the longest error correction and recovery time for $\tau$ (see Fig.~\ref{fig:FT_QEC_circuits}b and d), this analysis gives a lower bound on the memory lifetime, $\Gamma^{-1}$, limited by the $d-1$ coincident CREs within $\tau$
(see solid lines in Fig.~\ref{fig:rates}).

In contrast, for the non-FT implementation of the $[[7,1,3]]$ code, we obtain an upper bound on the memory lifetime
(dashed line in Fig.~\ref{fig:rates}). In this case, some double events cause a  logical failure. Since we are interested in an upper bound, we only consider the case  where the first erasure error occurs on the edge chips in Fig.~\ref{fig:FT_QEC_circuits}a. Following this event, depending on the affected chip, there are one or two stabilizer measurements during which an ancilla erasure can lead to logical failure. Therefore, for an upper bound, we consider CREs on these 6 edge chips with the rate $6\lambda$ as triggering events and find the probability of an additional event on an ancilla during one of the stabilizer measurements. Since different stabilizer measurements (colored boxes in Fig.~\ref{fig:FT_QEC_circuits}c) have the same number of inter-chip gates, we assume that they each take $\tau/6$. Therefore, we find the upper bound of $6\lambda[1-\exp(-\lambda \tau/6)]$ for $\Gamma$ in this case, see Fig.~\ref{fig:rates}.

Since the improvement sensitively depends on the recovery time, $\tau$, it is crucial to estimate the feasible recovery time for realistic superconducting devices.

\begin{figure}[t]
    \centering
    \includegraphics[width=0.95\columnwidth]{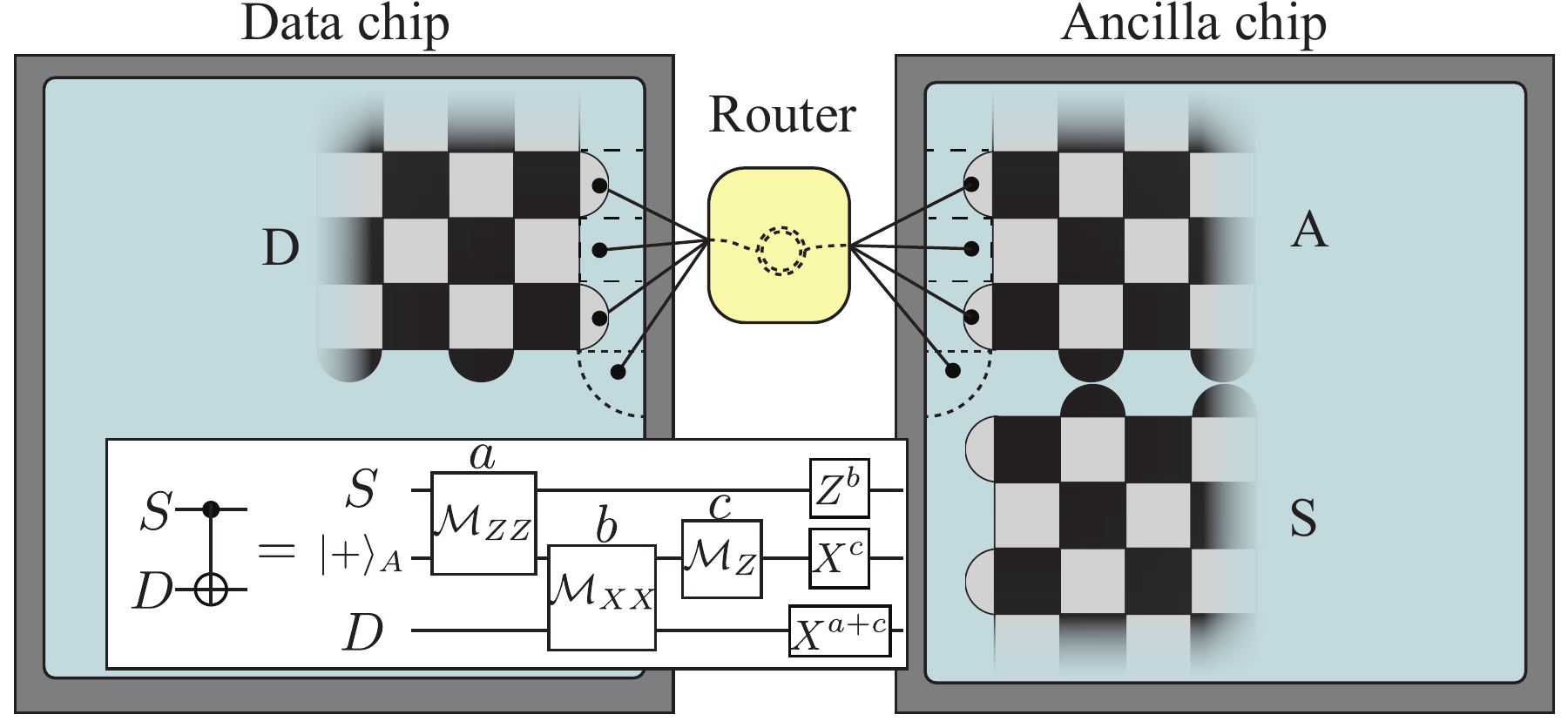}
    \caption{Experimental layout. A zooming-in on Fig.~\ref{fig:schematic}(a) that shows how the ancilla chip is coupled to a data chip. We show three surface patches S, A, and D, each encoded in a rotated surface code, with data qubits on the vertices and X-type (Z-type) syndrome qubits on the black (white) plaquettes. A $CX$ gate between the syndrome patch S and the data patch D, which is used for the syndrome extraction during the erasure QEC, is implemented by introducing the ancilla patch A and applying the measurement-based circuit shown in the inset at the lower-left corner. The dashed boxes indicate the new plaquettes to be measured while merging the boundaries between D and A nonlocally. The syndrome qubits associated with the new (dashed) plaquettes are drawn in black dots explicitly and they are coupled through the router via the microwave communication links indicated by the solid black lines.}
    \label{fig:exp_layout}
\end{figure}

{\it Experimental implementation.-- } Our proposed scheme can be implemented experimentally in superconducting devices by coupling multiple data chips to an ancilla chip through a router~\cite{Chang2020,zhou2021modular} as schematically illustrated in Fig.~\ref{fig:schematic}a. The ancilla chip is used to collect the syndrome information by coupling a syndrome patch to the data patches (all encoded in a surface code) associated with different stabilizers. We zoom in on Fig.~\ref{fig:schematic}a and show in detail how the ancilla chip is coupled to one of the data chips in Fig.~\ref{fig:exp_layout}. To implement an entangling gate, e.g. $CX$, between the syndrome patch S and the data patch D, we introduce an ancilla patch A on the ancilla chip and apply the measurement-based $CX$ gate~\cite{horsman2012surface}, whose circuit is shown in the inset at the lower-left corner of Fig.~\ref{fig:exp_layout}.
The measurement of joint Pauli operators $ZZ$ ($XX$) between the surface patches A and S (D) is implemented by lattice surgery~\cite{horsman2012surface}, i.e. merging and then splitting the $Z$ ($X$) boundaries of the two involved patches. The whole $CX$ gate using the lattice surgery is fault-tolerant in the surface-code level and compliant with the local constraints in 2D architecture~\cite{horsman2012surface}.
The nontrivial part of our setting is that we need to nonlocally merge the boundaries of the A and D patches that sit on different chips. This is done by adding new plaquettes (see the dashed boxes in Fig.~\ref{fig:exp_layout}) that connect the two nonlocal boundaries. For each of the new plaquette, we have two ancilla qubits (see the black dots in Fig.~\ref{fig:exp_layout}), each sitting on one chip and is locally coupled to two data qubits on the boundary of the surface patches. To measure the stabilizer associated with a new plaquette, we first apply a nonlocal $CX$ gate between the two ancilla qubits to create a Bell state $\frac{1}{\sqrt{2}}(|00\rangle + |11\rangle)$, then apply two $CX/CZ$ gates between the ancilla qubits and their coupled data qubits, apply another nonlocal CX gate between the ancillas and finally measure one of the ancillas. The non-local physical $CX$ gate between the ancilla qubits can be implemented by teleportation-based gates that use pre-shared and purified bell pairs between two chips as resources~\cite{Bennett1996,Wan2019,chou2018deterministic,yan2022entanglement}. 

We can estimate the length of the outer QEC cycle and the corresponding upper bound of the logical error rate based on realistic experimental parameters in the superconducting architecture. The most time-consuming physical operations are the two-qubit gates ($\sim 100\ $ns~\cite{Arute2019,Wu2021}), measurements ($\sim 200\ $ns~\cite{Heinsoo2018}) and inter-chip state transfers ($\sim 100\ $ns~\cite{Gold_2021,zhong2021deterministic}). We assume that each surface patch is a $10\times 10$ rotated surface code and each surface-level operation is followed by a full surface-QEC cycle with 10 rounds of repeated syndrome measurements. For maximum parallelism for all the operations, we estimate that the maximum recovery time $\tau$ for the [[4,1,2]] ([[7,1,3]]) code correcting 1 (2) erasure errors is approximately $270\ \mu$s ($1000\ \mu$s). See Supplementary Material~\cite{SM} for details. Therefore, based on these estimated recovery times,  we obtain a lower bound of the memory lifetimes of approximately 5 hours using the four-qubit, and 51 days for seven-qubit codes, see markers in Fig.~\ref{fig:rates}.

{\it Discussion.-- } So far we have focused on quantum memory and showed that we can protect the quantum system from catastrophic events for a sufficiently long time using distributed FTQEC. In principle, our scheme can be extended to universal fault-tolerant computing since it is compatible with the existing protocols. Furthermore, the resource overhead required for overcoming the CREs could be less than that required for the standard depolarizing noise.
For example, we can prepare the magic states non-fault-tolerantly and verify them by performing erasure detection, without applying costly magic-state distillation~\cite{knill2004fault, bravyi2005universal}, if we only aim to correct the CRE-induced erasure errors.

We can also use Knill-type QEC~\cite{knill2005scalable} to correct erasure errors. The Knill-QEC performs the error correction while teleporting the information from the data block to one of the blocks in a prepared encoded Bell pair. The fault tolerance of the Knill-QEC for general erasure errors is analyzed in Ref.~\cite{knill2005scalable}. In Supplementary Material~\cite{SM}, we show the application of the Knill-QEC in our setting. Compared to the erasure-flag scheme, the Knill scheme could be faster since the syndromes are measured in parallel. However, it is more resource-demanding since it requires two extra blocks of qubits encoded as a Bell pair for each logical qubit. Moreover, the preparation of the Bell pair potentially requires a complex coupling structure between the data chips.

Lastly, we discuss the possibility of optimizing the outer QEC to correct both the erasure errors and the Pauli errors uncorrectable by the surface codes. For now, the introduction of the second layer of QEC exponentially suppresses the error rate $\gamma_L^{E}$ due to the rare erasure errors while linearly enhancing the error rate $\gamma_L^{P}$ due to the Pauli errors resulting from the failure of the surface codes. In the regime where $\gamma_L^{E} \ll \gamma_L^{P}$, it is advantageous to consider the tradeoff between $\gamma_L^{E}$ and $\gamma_L^{P}$ and minimize the total logical error rate $\gamma_L = \gamma_L^{E} + \gamma_L^{P}$ by tailoring the outer codes to correct both erasure and Pauli errors. However, the details of the code tailoring, the fault-tolerant QEC design, and the implementation of the universal gates, which might require magic state distillation, remain to be explored.

\begin{acknowledgments}
We thank Vignesh Raman for helpful discussions. 
We acknowledge support from the ARO (W911NF-18-1-0020, W911NF-18-1-0212), ARO MURI (W911NF-16-1-0349, W911NF-21-1-0325), AFOSR MURI (FA9550-19-1-0399, FA9550-21-1-0209), AFRL (FA8649-21-P-0781), DoE Q-NEXT, NSF (OMA-1936118, EEC-1941583, OMA-2137642), NTT Research, and the Packard Foundation (2020-71479). A.S. is supported by a Chicago Prize Postdoctoral Fellowship in Theoretical Quantum Science. BOS and RV acknowledge the support from JST Moonshot R$\&$D Grant (JPMJMS2061).
\end{acknowledgments}

\bibliographystyle{apsrev4-1}
% \bibliography{dis_qec}


\end{document}

%\preprint{}
%Title of paper$\epsilon_{G,m}^{(1)}(T)$
\title{Supplemental Information for ``Distributed quantum error correction for chip-level catastrophic errors"}
\author{Qian Xu} 
\thanks{These authors contributed equally.}
\affiliation{Pritzker School of Molecular Engineering, The University of Chicago, Chicago IL 60637, USA}

\author{Alireza Seif} 
\thanks{These authors contributed equally.}
\affiliation{Pritzker School of Molecular Engineering, The University of Chicago, Chicago IL 60637, USA}

\author{Haoxiong Yan}
%\affiliation{IQIM, California Institude of Technology, Pasadena, CA 91125, USA}
\affiliation{Pritzker School of Molecular Engineering, The University of Chicago, Chicago IL 60637, USA}

\author{Nam Mannucci}
%\affiliation{IQIM, California Institude of Technology, Pasadena, CA 91125, USA}
\affiliation{Pritzker School of Molecular Engineering, The University of Chicago, Chicago IL 60637, USA}

\author{Bernard~Ousmane~Sane}
\affiliation{Graduate School of Media and Governance, Keio University, 5322 Endo, Fujisawa 252-0882, Japan}

\author{Rodney Van Meter}
\affiliation{Faculty of Environment and Information Studies, Keio University, 5322 Endo, Fujisawa 252-0882, Japan}

\author{Andrew N. Cleland}
%\affiliation{IQIM, California Institude of Technology, Pasadena, CA 91125, USA}
\affiliation{Pritzker School of Molecular Engineering, The University of Chicago, Chicago IL 60637, USA}
\affiliation{Center for Molecular Engineering and Material Science Division, Argonne National Laboratory, Lemont IL 60439, USA}

\author{Liang Jiang} 
\affiliation{Pritzker School of Molecular Engineering, The University of Chicago, Chicago IL 60637, USA}
%\affiliation{AWS Center for Quantum Computing, Pasadena, CA 91125, USA}

\maketitle
\tableofcontents
\section{Fault-tolerant criteria for the erasure error correction}
First, we present the criteria that a fault-tolerant (FT) scheme should satisfy for correcting erasure errors, following the standard FT criteria for depolarizing noise~\cite{Aliferis2006QuantumAcc, gottesman2010introduction, chamberland2018flag}. For $t = d - 1$, an erasure QEC protocol is $t$ fault-tolerant using a distance-$d$ stabilizer code if the following two conditions are satisfied:
\begin{enumerate}
    \item For an input codeword with an error of weight $s_1$, if $s_2$ erasures occur during the protocol with $s_1 + s_2 \leq t$, ideally decoding the output state gives the same codeword as ideally decoding the input state.
    \item For $s$ erasures during the protocol with $s \leq t$, no matter how many errors are present in the input state, the output state differs from a codeword by an error of at most weight $s$.
\end{enumerate}
One can check that the erasure-flag scheme presented in the main text satisfies the above two FT criteria.

\section{The Knill quantum error correction for erasure errors}
\begin{figure}[h!]
    \centering
    \includegraphics[width = 1\textwidth]{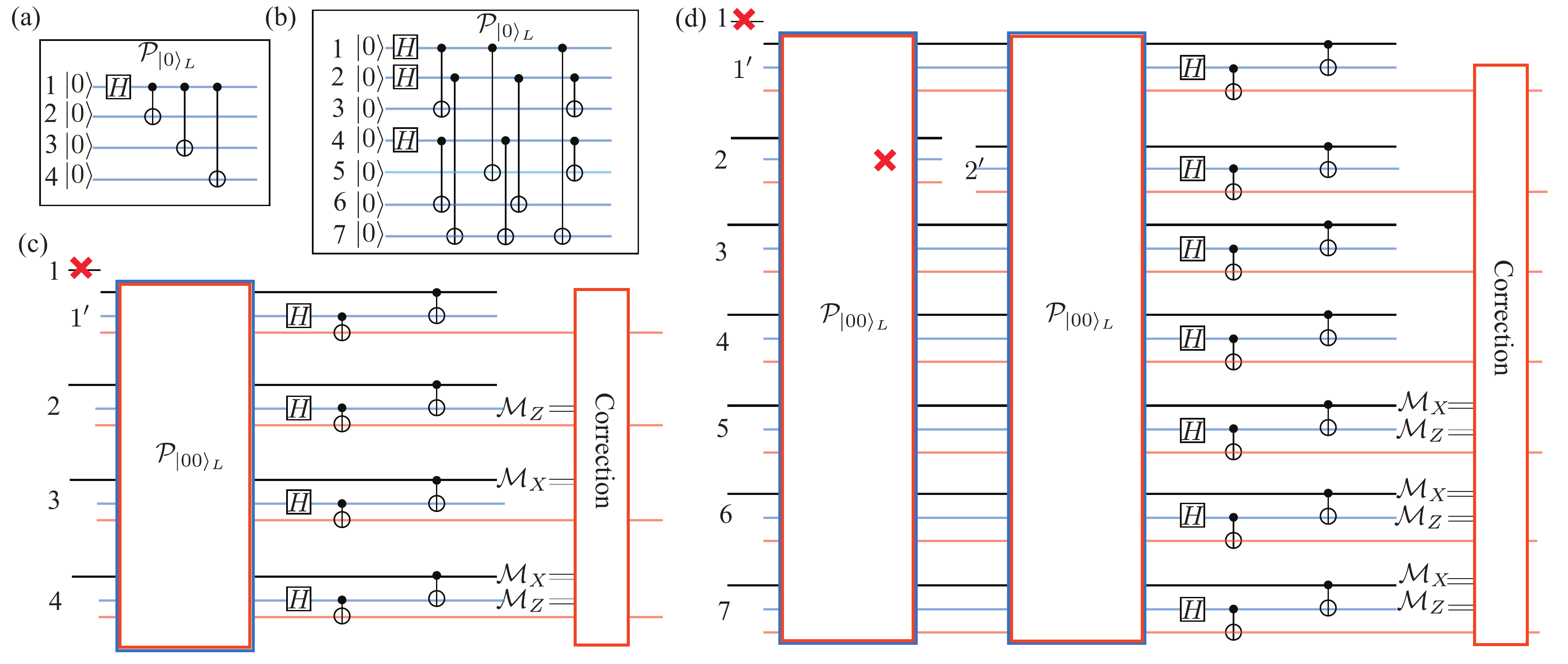}
    \caption{The Knill QEC circuits. The non-transversal circuits for the preparation of logical $|0\rangle$ for the (a) $[[4,1,2]]$ code, (b) $[[7,1,3]]$ code. (c). The FT circuit for the $[[4,1,2]]$ code correcting 1 data erasure at the input. (d). The FT circuit for the $[[7,1,3]]$ code correcting 1 data erasure at the input and 1 erasure during the preparation of $|00\rangle_L$. For (c) and (d), there are three blocks of qubits (black, blue and orange), each consisting of $n$ qubits encoded in an $n$-qubit code that are distributed over $n$ chips indexed by different numbers. An encoded bell pair between the blue and the orange blocks is created by preparing the logical state $|00\rangle_L$ non-transversally using the circuits (a)(b), and then applying a transversal (logical) Hadmard and a $CX$ gate. The information then is teleported from the black block to the orange block by performing a bell measurement between the black and the blue blocks. The bell measurement is implemented by first applying a transversal CX gate between the black and the blue blocks and measuring a logical $X$ ($Z$) operator on the black (blue) block, whose supported qubits have not been erased. The correction for the erasure errors is done by applying a Pauli frame update on the output orange block conditioned on the measurement outcomes.}
    \label{fig:knill_circuits}
\end{figure}

% \subsection{The Knill-QEC scheme}
In this section, we adapt the Knill-QEC scheme~\cite{knill2005scalable} to our setting to correct catastrophic erasure errors on different chips. For a logical qubit encoded in an $n$-qubit code, we use three blocks of qubits indexed by different colors (black, blue, and orange). Each block is encoded using the $n$-qubit code, with each qubit in a distinct chip.
%with $n$ qubits distributed over $n$ chips.
In other words, there are $n$ chips in total, each containing three qubits of different colors (see Fig.~\ref{fig:knill_circuits}(c) and (d)). To correct the erasure errors, we first prepare an encoded Bell pair between the blue and orange blocks and then teleport the information from the black block to the orange block by performing a Bell measurement on the black and blue blocks. The erasure-QEC succeeds if the logical information is not erased. That is, if we can express the logical $X$ and $Z$  operators using the remaining unaffected qubits. Faithful measurements of these logical $X$ and $Z$ operators recover the information in the black and blue blocks, respectively.
Although most of the required operations are transversal, i.e. no inter-chip gates are required, the preparation of the logical $|00\rangle_L$ state for the blue and orange blocks is non-transversal (see Fig.~\ref{fig:knill_circuits}(a)(b)). Consequently, an erasure error that happens in the middle of the logical-zero preparation could spread among different chips and erase multiple qubits in the blue/orange blocks. Fortunately, this problem can be resolved by verifying the state and redoing the preparation if necessary. Specifically, we can fault-tolerantly operate an $[[n,k,d]]$ code against erasure errors by implementing the following protocol:  

(i) If there are data erasures detected at the beginning, replace the erased data qubits (chips) and apply the following erasure-QEC. (ii) Use the non-transversal circuit to prepare the logical state $|0_L\rangle$ for both the blue and the orange blocks. If there are bad erasures detected in the middle, replace the erased qubits (chips) and restart the state preparation. Repeat until the logical $|0 0\rangle_L$ is prepared without bad erasures. Then apply transversal gates to prepare the bell state $|\Phi^+_L\rangle$ (between the blue and orange blocks). Next, perform the transversal Bell measurement between the black block and the blue block by measuring the intact qubits (which have not been erased) and teleport the information from the black block to the orange block. 

As examples, we explicitly draw the fault-tolerant circuits for the $[[4,1,2]]$ ($[[7,1,3]]$) code correcting 1 (2) erasure errors in Fig.~\ref{fig:knill_circuits}(c) (Fig.~\ref{fig:knill_circuits}(d)). We note that the circuit for the $[[7,1,3]]$ is adaptive and we need to redo the preparation of $|00\rangle_L$ if there is erasure detected in the middle.

\section{Comparison between the Knill scheme and the erasure-flag scheme}
We provide a comprehensive comparison between the Knill and erasure-flag schemes in terms of their resource overhead, implementation speed, and requirements for the hardware layout. In general, the erasure-flag scheme is more resource-efficient, yet more complex and slower due to longer sequential circuits. The Knill-QEC, on the other hand, is simpler, faster, yet more resource-demanding and potentially requiring complex connectivity between different chips.

We first compare the resource overhead. For the $[[4,1,2]]$ code, the Knill-scheme requires 12 surface patches, whereas the erasure-flag scheme only requires 5 patches. In general, for an $[[n,k,d]]$ code, the Knill-scheme requires $3n$ patches, while the erasure-flag requires $n + 1$ patches. If we take into account the ancilla patches used for lattice surgery, in the general case, the Knill-scheme requires $4n$ patches, while the erasure-flag requires $n + 2$ patches.)

To compare the time requirement of these schemes, we roughly estimate the maximal cycle correcting 1 (2) erasure errors for the four-(seven-) qubit codes using the two schemes. As shown in the main text, we take into account the following most time-consuming operations in the physical level for the superconducting architecture: two-qubit gates ($\sim 100\ $ns), measurements ($\sim 200\ $ns) and inter-chip state transfers ($\sim 100\ $ns). An inter-chip $CX/CZ$ gate using pre-shared and purified bell pair~\cite{chou2018deterministic}, which mainly consists of one step of local entangling gates and one step of measurement, takes about $t_{\textrm{NCX}} \approx 300 $ ns. We then estimate the time for the operations at the surface-code level. We denote the time for a full surface-code QEC cycle as $t_{\textrm{SQ}}$. A full cycle consists of 10 rounds of repeated stabilizer measurements (for a $10 \times 10$ rotated surface code), each consisting of four steps of parallel $CX$ ($CZ$) gates between the syndrome qubits and the data qubits and a measurement on the syndrome qubits. Therefore, $t_{\textrm{SQ}} \approx 10\times (4\times 100\ {\rm{ns}} + 200\ {\rm{ns}}) \approx 6\ \mu$s. The state preparation, measurement and single-qubit operations in the surface-code level take roughly $t_{\textrm{SQ}}$, while the two-qubit gates based on lattice surgery are more time-consuming. A local measurement-based $CX/CZ$ gate, which is introduced in Ref.~\cite{horsman2012surface} (see also Fig.~4 in the main text), mainly consists of one step of state preparation, two steps of joint Pauli measurement between two surface patches and one step of single-patch measurement. Each joint measurement between two surface patches is implemented by a merging and a splitting of the two patches, each followed by a full surface QEC cycle~\cite{horsman2012surface}. So in combination a local surface $CX/CZ$ gate takes about $t_{\textrm{SCX}} \approx 6 t_{\textrm{SQ}} \approx 36\ \mu$s. An inter-chip $CX/CZ$ gate at the surface-code level is implemented similarly, except that the merging between two surface patches on two different chips (the ancilla patch and the data patch in Fig.~4 in the main text) takes extra time due to the implementation of inter-chip physical $CX$ gates. As such, each inter-chip surface $CX/CZ$ gate takes a time $t_{\textrm{NSCX}} \approx t_{\textrm{SCX}} + 10 t_{\textrm{NCX}} \approx 39\ \mu$s. Given the estimation of each surface-level operation, we can proceed to estimate the maximal cycle duration for the four- and seven-qubit codes, using the erasure-flag and the Knill circuits shown in Fig.~2 in the main text and Fig.~\ref{fig:knill_circuits}, respectively. The estimates are summarized in Tab.~\ref{tab:runtime_estimation}. We note that these rough estimates strongly depend on the schemes and assumptions for different experimental operations, which are by no means optimal now. The Knill scheme is in general faster than the erasure-flag scheme since the latter requires more sequential operations using only one ancilla surface patch. However, we note that if we use more ancilla surface patches and measure different stabilizers in parallel, the erasure-flag scheme can be greatly accelerated.

In terms of the experimental layout, the erasure-flag scheme is preferred since it only requires the connectivity between the ancilla chip and the data chips, whereas the Knill-scheme might require complex connectivity between the data chips while preparing the logical state $|0\rangle_L$ (see Fig.~\ref{fig:knill_circuits}(b)).

\begin{table}[t]
    \centering
    \begin{tabular}{p{2cm}<{\centering}|p{6cm}<{\centering}|p{6cm}<{\centering}}
         \hline
          & $[[4,1,2]]$ & $[[7,1,3]]$ \\
         \hline
          Erasure-flag & $5 t_{\textrm{SQ}} + 6 t_{\textrm{NSCX}} \approx 270 \mu$s &  $13 t_{\textrm{SQ}} + 24 t_{\textrm{NSCX}} \approx 1000 \mu$s \\
         \hline
         Knill & $4 t_{\textrm{SQ}} + 2 t_{\textrm{SCX}} + 3 t_{\textrm{NSCX}} \approx 210 \mu$s &  $6 t_{\textrm{SQ}} + 2 t_{\textrm{SCX}} + 6 t_{\textrm{NSCX}} \approx 340 \mu$s \\
         \hline
    \end{tabular}
    \caption{Estimation of the maximal recovery time of a QEC cycle correcting 1 (2) erasure errors for the 4 (7)-qubit codes based on experimentally-relevant parameters. Each erasure-flag circuit shown in Fig.~2 in the main text consists of one round of erasure detection (surface QEC) at the beginning, which takes $t_{\textrm{SQ   }}$, and the measurements of multiple stabilizers. A measurement of a weight-$w$ stabilizer consists of the initialization of the ancilla qubit, $w$ sequential inter-chip CX/CZ gates between the ancilla qubit and the data qubits, and a measurement on the ancilla. So each weight-$w$ stabilizer measurement takes $T_w \approx 2t_{\textrm{SQ}} + w t_{\textrm{NSCX}}$. As shown in Fig.~2(b) in the main text, the worst-case erasure-flag circuit for the $[[4,1,2]]$ code correcting 1 erasure consists of one round of initial erasure detection, one weight-4 stabilizer measurement and one weight-2 stabilizer measurement. As such, it takes roughly $t_{\textrm{SQ}} + T_4  + T_2 =  5 t_{\textrm{SQ}} + 6 t_{\textrm{NSCX}} \approx 270 \mu$s. As shown in Fig.~2(d) in the main text, the worst-case FT erasure-flag circuit for the $[[7,1,3]]$ code correcting 2 erasures consists of one round of initial erasure detection and six weight-4 stabilizer measurements. As such, it takes roughly $t_{\textrm{SQ}} + 6 T_4 = 13 t_{\textrm{SQ}} + 24 t_{\textrm{NSCX}} \approx 1000 \mu$s. On the other hand, as shown in Fig.~\ref{fig:knill_circuits}(c)(d), each Knill circuit consists of one round of initial erasure detection ($t_{\textrm{SQ}}$), possibly multiple rounds of logical state preparation $\mathcal{P}_{|00\rangle_L}$ ($T_{00}$), one round of Hadamard gates and two rounds of local CX gates ($t_{\textrm{SQ}} + 2t_{\textrm{SCX}}$). The preparation of logical state $|00\rangle_L$ is the most time-consuming part due to the sequential implementation of multiple inter-chip $CX$ gates. For the $[[4,1,2]]$ code, $T_{00} = 2 t_{\textrm{SQ}} + 3 t_{\textrm{NSCX}}$, and the maximal QEC cycle shown in Fig.~\ref{fig:knill_circuits}(c) takes roughly $2 t_{\textrm{SQ}} + 2t_{\textrm{SCX}} 
    + T_{00} = 4 t_{\textrm{SQ}} + 2t_{\textrm{SCX}} + 3 t_{\textrm{NSCX}} \approx 210 \mu$s. For the [[7,1,3]] code, $T_{00} = 2 t_{\textrm{SQ}} + 3 t_{\textrm{NSCX}}$, and the maximal QEC cycle shown in Fig.~\ref{fig:knill_circuits}(d) takes roughly $2 t_{\textrm{SQ}} + 2t_{\textrm{SCX}} 
    + 2 T_{00} = 6 t_{\textrm{SQ}} + 2t_{\textrm{SCX}} + 6 t_{\textrm{NSCX}} \approx 342 \mu$s.}
    \label{tab:runtime_estimation}
\end{table}
%\end{center}

\section{Application of the erasure-flag scheme to other codes}
In this section, we show that the erasure-flag scheme can in principle be applied to other codes in addition to the $[[4,1,2]]$ and $[[7,1,3]]$ codes shown in the main text. 

As an example, we show that it can be applied to the CSS surface codes, or more generally, Kitaev toric codes~\cite{kitaev2003fault} with an arbitrary distance. We consider a $d\times d$ planar surface code as an example and prove that by using the erasure-flag scheme, the surface code can correct up to $d - 1$ arbitrary erasures. The proof for other toric codes follows. The key is to show that by adaptively measuring an appropriate set of stabilizers in a proper sequence, the erasure-flag error set never grows to be too large to be uncorrectable as long as there are no more than $d - 1$ erasures during the protocol. We equivalently view an erasure error as an arbitrary Pauli error with a known location. 

To correct the possible errors in an erasure-flag error set $\mathcal{E}$, we measure only a minimal set of stabilizers that are sufficient to diagnose and correct the errors. That is, for $s$ possible $X$ errors on $s$ known qubits, we measure at most $s$ $Z$-type stabilizers whose support cover the $s$ faulty qubits. The same goes for the $Z$ errors. Suppose that a logical $X$ operator has to vertically cross the surface patch and has support on $d$ different rows. Let $s_i$ be the total number of erasures that have occurred and $r_i$ be the total number of rows on which the current erasure-flag error set $\mathcal{E}$ have support  before the $i$th stabilizer measurement. We have the iterative relation $r_{i + 1} - r_i \leq s_{i + 1} - s_i$. To see this, suppose one or more bad erasures happen and propagates to multiple data errors while measuring a new stabilizer that is supported on at least one of the rows which $\mathcal{E}$ also have support on. Since a stabilizer only spans two rows, the expanded error set $\mathcal{E}$ only occupies one more row. As a result, we have $r_i \leq s_i$, and for $s_i \leq d - 1$ we have $r_i \leq d - 1$. Therefore, the product of any two errors in  $\mathcal{E}$ does not form a logical $X$ operator provided that there are no more than $d - 1$ erasures in total. An analogous argument show that the flag error set can not fully span the $d$ columns and the product of two errors can not be a logical $Z$ operator.

\section{System Performance Implications}

The previous proposals for distributed hardware discussed in the main text have focused on achieving scalability by increasing the number of logical qubits that can be deployed on a single computation. In such systems, for a broad range of parameters, the application execution time is determined in large part by the inter-node entanglement creation time, interconnect topology and scheduling application-matched use of the interconnect~\cite{van-meter06:thesis,ahsan2015designing}.

In contrast, in this proposal, the distributed hardware is deployed to mitigate this CRE problem.  In the simplest model for system organization, each of the $n$ nodes has enough logical qubits to run the entire application. To first approximation, the distributed nature of the system has no impact on application execution time. However, recovery operations after CREs will occupy individual logical qubits within the nodes for extended periods of time, as each logical qubit must execute inter-node gates by using physical Bell pairs created using the dedicated transceiver qubits that exist in specific locations on the chip.

The optimal mapping of qubits to locations on a chip, taking into account the intra-chip communication costs, is known to be an NP-complete problem for compilation of static programs~\cite{Siraichi:2018:QA:3179541.3168822}. The scheduling of the use of the physical Bell pairs for inter-chip recovery can impact the preferred location of logical qubits within a chip and delay the start of forthcoming application logical gates. The details of qubit movement and the need to dynamically schedule this recovery process to minimize its impact on application performance is an important area for future research.

The hardware cost for this proposal is a factor of six (nine) when using the [[4,1,2]] ([[7,1,3]]) code, including the data code word nodes, the ancilla (outer QEC syndrome) node, and at least one node held in reserve for recovery operations, plus the interconnect itself. Thus, the hardware cost is substantial. However, the gain in projected memory lifetime is four to six orders of magnitude, an enormous improvement. For long computations, this gain is absolutely imperative and will determine what computations can and cannot be performed using such solid-state quantum computers.

% % \input{SupplementaryMat.bbl}
% \bibliography{SM} 
\bibliographystyle{apsrev4-2}